\documentclass[aip,jcp,amsmath,reprint]{revtex4-2}

\usepackage[utf8]{inputenc}
\usepackage{graphicx}
\usepackage{hyperref}
\usepackage{amsmath,amssymb}
\usepackage[capitalize]{cleveref}
\usepackage{bbm}
\usepackage{algorithm}
\usepackage{algpseudocode}
\crefname{section}{Sec.}{Sec.}
\crefname{figure}{Fig.}{Fig.}
\renewcommand{\vec}[1]{\mathbf{#1}}

\usepackage{color}

\begin{document}

\title{Solid-Angle Nearest-Neighbor Method for Size-Disperse Systems of Spheres}
\author{Nydia Roxana Varela-Rosales}
\email{nydia@fun.ac.jp}
\affiliation{Institute for Multiscale Simulation, IZNF, Friedrich-Alexander-Universität Erlangen-Nürnberg, 91058 Erlangen, Germany}
\affiliation{Department of Complex and Intelligent Systems, Future University Hakodate, 041-8655, Hokkaido, Japan}
\author{Michael Engel}
\email{michael.engel@fau.de}
\affiliation{Institute for Multiscale Simulation, IZNF, Friedrich-Alexander-Universität Erlangen-Nürnberg, 91058 Erlangen, Germany}

\date{\today}

\begin{abstract}
Identifying nearest neighbors accurately is essential in particle-based simulations, from analyzing local structure to detecting phase transitions. While parameter-free methods such as Voronoi tessellation and the solid-angle nearest-neighbor (SANN) algorithm are effective in monodisperse systems, they become less reliable in mixtures with large size disparities. We introduce SANNR, a generalization of SANN that incorporates particle radii into the solid-angle criterion for robust, size-sensitive neighbor detection. We compare SANNR against Voronoi, Laguerre, and SANN in binary and size-disperse sphere mixtures. Using Wasserstein distance metrics, we show that SANNR closely matches size-aware Laguerre tessellation while preserving the geometric continuity of SANN. Applied to the crystallization of the complex AB$_{13}$ phase, SANNR improves detection of local bond-orientational order and better captures the emergence of global symmetry. SANNR thus offers a smooth, parameter-free, and extensible framework for neighbor detection in polydisperse and multicomponent systems.
\end{abstract}

\maketitle

\section{Introduction}

Characterizing the local environment of particles is central to the analysis of structural order in atomic, molecular, colloidal, and granular systems. Many theoretical and computational methods rely on geometric descriptors to define short-range correlations, quantify local order, or detect structural motifs. Among these, identifying the nearest-neighbor shell, the set of particles surrounding a reference particle, is a foundational step. It forms the basis for computing bond orientational order parameters,\cite{Steinhardt1983,Lechner2008,Stukowski2012,Mickel2013,Eslami2018,Haeberle2019} detecting local symmetries,\cite{Malins2013,LazarE5769,Morse2016} and analyzing phase transitions and crystallization processes.

A widely used method for nearest-neighbor identification is Voronoi tessellation, which partitions space into cells such that each point belongs to the region closest to a specific particle.\cite{Voronoi1908} Two particles are considered neighbors if their corresponding Voronoi cells share a face. While Voronoi tessellation is conceptually elegant and parameter-free, it treats all particles as point-like and thus fails to account for size disparities. In systems with large differences in particle size, such as colloidal suspensions or binary alloys,\cite{Ohara1995,Cabane2016} the resulting neighbor lists may be physically inaccurate: for example, cell faces may cut into larger particles,\cite{Richard1998} or close-packed configurations may yield distorted neighbor networks. Furthermore, Voronoi tessellation becomes increasingly complex and computationally expensive in higher dimensions or large-scale simulations.

To address these limitations, Laguerre tessellation (also known as radical or power tessellation) introduces radius-dependent weighting into the Voronoi construction.\cite{Aurenhammer1987,Okabe2009} This approach shifts cell boundaries to reflect particle volume, providing more physically meaningful neighbor relations in size-disperse systems. Laguerre tessellation has proven effective in analyzing binary\cite{Richard1998,Richard2001} and polydisperse mixtures,\cite{Travesst2017} but remains sensitive to thermal fluctuations and is computationally intensive. Small displacements or perturbations can cause discontinuous changes in the neighbor network, particularly in disordered environments like glasses or liquids.\cite{Medvedev1988}

An alternative approach is the Solid Angle Nearest Neighbor (SANN) algorithm,\cite{vanMeel2012} which defines neighbors by summing the solid angles subtended by surrounding particles until a full sphere ($4\pi$ steradians) is covered. This method avoids explicit partitions of space and instead focuses on local geometry around each particle. SANN is efficient, robust to noise, and produces stable neighbor lists under thermal motion.\cite{Staub2020,Mugita2024} However, like Voronoi tessellation, SANN is blind to particle size and is therefore limited when applied to size-disperse systems.

To overcome this, we introduce the Regularized Solid Angle Nearest Neighbor (SANNR) algorithm, a generalization of SANN that incorporates particle radii when calculating the blocked solid angle.\cite{Mugita2024} SANNR retains the computational efficiency and robustness of the original method while introducing a size-aware regularization scheme. It smoothly interpolates between radius-independent and radius-sensitive neighbor assignments, making it well suited for complex fluids, polydisperse colloids, and systems with strong size asymmetry.

In this work, we develop and formalize the SANNR algorithm, compare it to Voronoi, Laguerre, and SANN methods in both binary and polydisperse systems, and evaluate its effectiveness for characterizing local structure via bond-orientational order parameters. We demonstrate that SANNR not only produces neighbor distributions closer to those of Laguerre tessellation, but also improves the sensitivity and accuracy of structural diagnostics in systems undergoing crystallization.

\section{Nearest-Neighbor Determination}

An effective method for identifying nearest neighbors should meet three essential criteria: (1)~it should be free of tunable parameters and rely solely on geometric information, (2)~it must remain accurate across a wide range of particle densities, and (3)~it should correctly capture local structure in systems with phase coexistence. While using a fixed-distance cutoff is simple, this approach fails to satisfy these criteria and often performs poorly in heterogeneous or disordered systems. Consequently, parameter-free geometric methods have become essential tools for reliable nearest-neighbor identification.

\subsection{Voronoi and Laguerre Tessellation}\label{Voronoi_and_Laguerre_intro}

Geometric tessellation methods partition space into convex polyhedral cells, using the spatial arrangement of particle centers. In the standard Voronoi tessellation, the cell $V_i$ associated with particle $P_i$ is defined as
\begin{equation}\label{eq:VoronoiDefinition}
	V_i = \{\vec{x} \in \mathbbm{R}^3 \mid d(\vec{x}, \vec{p}_i) \leq d(\vec{x}, \vec{p}_j)\; \text{for all}\; j \neq i\},
\end{equation}
where $d$ is the Euclidean distance and $\vec{p}_i$ is the position of $P_i$. In three dimensions, this yields convex polyhedra whose faces separate points equidistant to neighboring particles. Two particles are considered neighbors if their cells share a face.

\begin{figure}
    \includegraphics[width=\columnwidth]{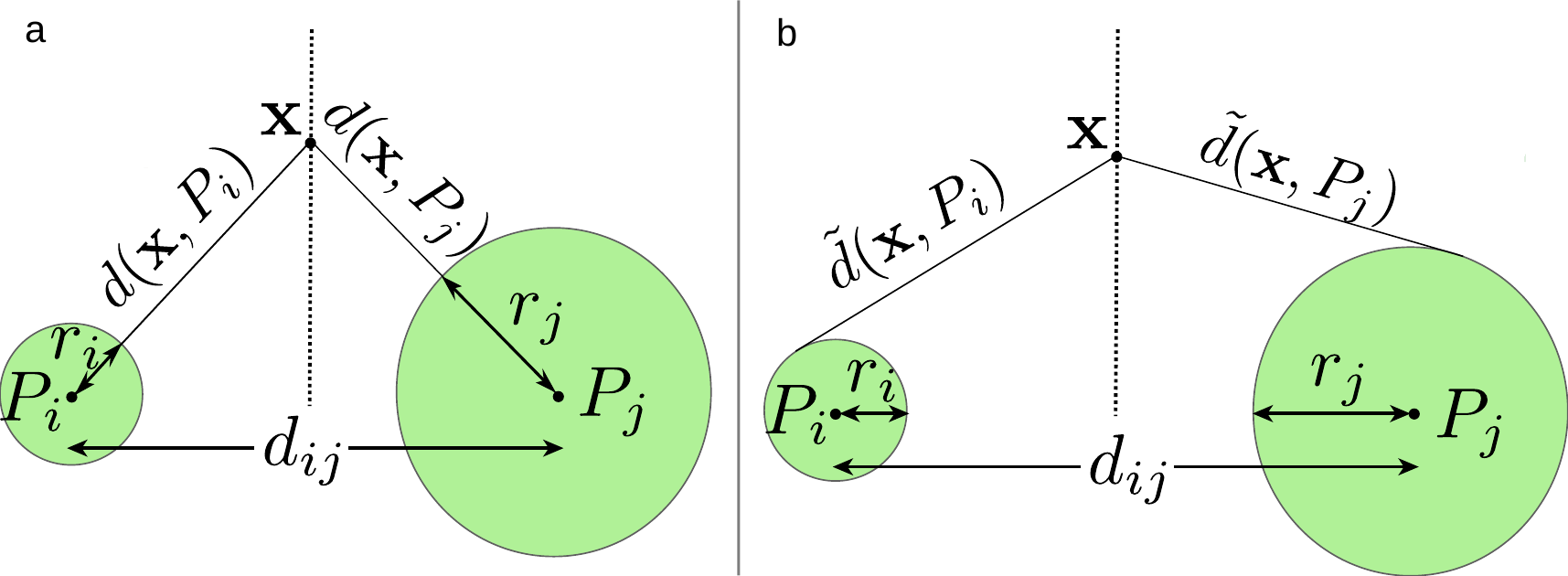}
	\caption{Construction of cell faces between spherical particles $P_i$ and $P_j$. \textbf{a},~Voronoi tessellation places the face equidistant from centers using the Euclidean distance $d$, often leading to the boundary intruding into larger particles. \textbf{b},~Laguerre tessellation uses the power distance $\tilde{d}$, shifting the face toward the smaller particle and better capturing excluded volume.}
	\label{fig.Voronois}
\end{figure}

However, standard Voronoi tessellation does not account for particle radii, making it unreliable in systems with size disparity. As shown in \cref{fig.Voronois}a, the Voronoi boundary may intrude into the volume of larger particles, even when particles are non-overlapping, leading to unphysical neighbor assignments.

To address this issue, the Laguerre tessellation (also known as power or radical tessellation)\cite{Aurenhammer1987,Okabe2009} incorporates particle size via a modified distance:
\begin{equation}\label{eq:LaguerreVoronoiDefinition}
	V_i = \{\vec{x} \in \mathbbm{R}^3 \mid \tilde{d}(\vec{x}, P_i) \leq \tilde{d}(\vec{x}, P_j)\; \text{for all}\; j \neq i\},
\end{equation}
where the power distance is defined as
\begin{equation*}
\tilde{d}(\vec{x}, P_i) = \sqrt{d(\vec{x}, \vec{p}_i)^2 - r_i^2},
\end{equation*}
and $r_i$ is the radius of particle $P_i$.\cite{Gavrilova1996} This radius-aware formulation ensures that boundaries reflect excluded volume. As illustrated in \cref{fig.Voronois}b, the separating plane between particles shifts toward the smaller one, resolving the overlap issue and yielding more meaningful neighbor definitions in size-disperse systems.

\subsection{SANN Algorithm}\label{SANN_intro}

The solid-angle nearest-neighbor (SANN) algorithm~\cite{vanMeel2012} identifies neighbors based on angular coverage rather than distance thresholds or geometric partitions. For a given particle $P_i$, other particles are sorted by increasing distance $d_{i,j} = d(\vec{p}_i, \vec{p}_j)$. Each candidate neighbor subtends a solid angle at $P_i$ given by
\begin{equation}\label{eq:solidAngle}
	\Omega_{i,j} = 2\pi(1 - \cos{\theta_{i,j}}),
\end{equation}
where the cosine of the cone half-angle is $\cos(\theta_{i,j}) = d_{i,j} / R_i^{N_c}$, and $R_i^{N_c}$ is the cutoff radius associated with $N_c$ neighbors.

The algorithm determines $N_c$ and $R_i^{N_c}$ by requiring that the sum of solid angles blocked by the $N_c$ closest neighbors equals the full solid angle of a sphere:
\begin{equation}
	\sum_{j=1}^{N_c} \Omega_{i,j} = 4\pi.
\end{equation}
Solving yields the cutoff radius
\begin{equation}\label{eq:SANNdefinition}
	R_i^{N_c} = \frac{\sum_{j=1}^{N_c} d_{i,j}}{N_c - 2}.
\end{equation}
This equation is evaluated iteratively, increasing $N_c$ until the consistency condition
\begin{equation}\label{eq:SANNconstraint}
	d_{i,N_c} \leq R_i^{N_c} < d_{i,N_c+1}
\end{equation}
is satisfied. This guarantees that exactly $N_c$ particles lie within the calculated cutoff radius.

SANN is fully parameter-free, robust to density fluctuations, and particularly effective in disordered or dynamic systems where fixed cutoffs may fail.

\subsection{SANNR: Extension to Size-Disperse Systems}\label{SANNR_intro}

While SANN is geometry-based and efficient, it does not account for particle radii. To overcome this, we generalize the method to size-disperse systems via the regularized SANN (SANNR) algorithm, which incorporates particle sizes into the solid-angle calculation.

We modify the blocking angle to include radius effects:
\begin{equation}\label{eq:SANNregBlocking}
	\tan(\theta_{i,j}) = \frac{r_j + R_s}{d_{i,j}},
\end{equation}
where $r_j$ is the radius of neighbor $P_j$ and $R_s$ is a local regularization parameter determined from the nearest geometry.

To compute $R_s$, we first define
\begin{equation*}
	d_i = d_{i,1} + d_{i,2} + d_{i,3},
\end{equation*}
the cumulative distance to the three closest neighbors, which ensures angular coverage is well-defined in three dimensions. The total effective radius is given by
\begin{equation}
	R_{i,j}^{\text{tot}} = \sqrt{d_i^2 - \left(d_{i,j} - (r_i - r_j)\right)^2},
\end{equation}
and the blocking contribution is then regularized by
\begin{equation}
	R_s = R_{i,j}^{\text{tot}} - r_j.
\end{equation}

Substituting this into the solid-angle condition yields the generalized neighbor constraint:
\begin{equation}\label{eq:SANNregStart}
    N_c - 2 = \sum_{j=1}^{N_c} \frac{1}{\sqrt{1 + \left( \frac{r_j + R_s}{d_{i,j}} \right)^2 }},
\end{equation}
leading to the SANNR cutoff radius:
\begin{equation}\label{eq:SANNmainEquation1}
    R_i^{N_c} = \frac{\sum_{j=1}^{N_c} d_{i,j}}{\sum_{j=1}^{N_c} \left( 1 + \left( \frac{r_j + R_s}{d_{i,j}} \right)^2 \right)^{-1/2}}.
\end{equation}

\begin{figure}
\includegraphics[width=\columnwidth]{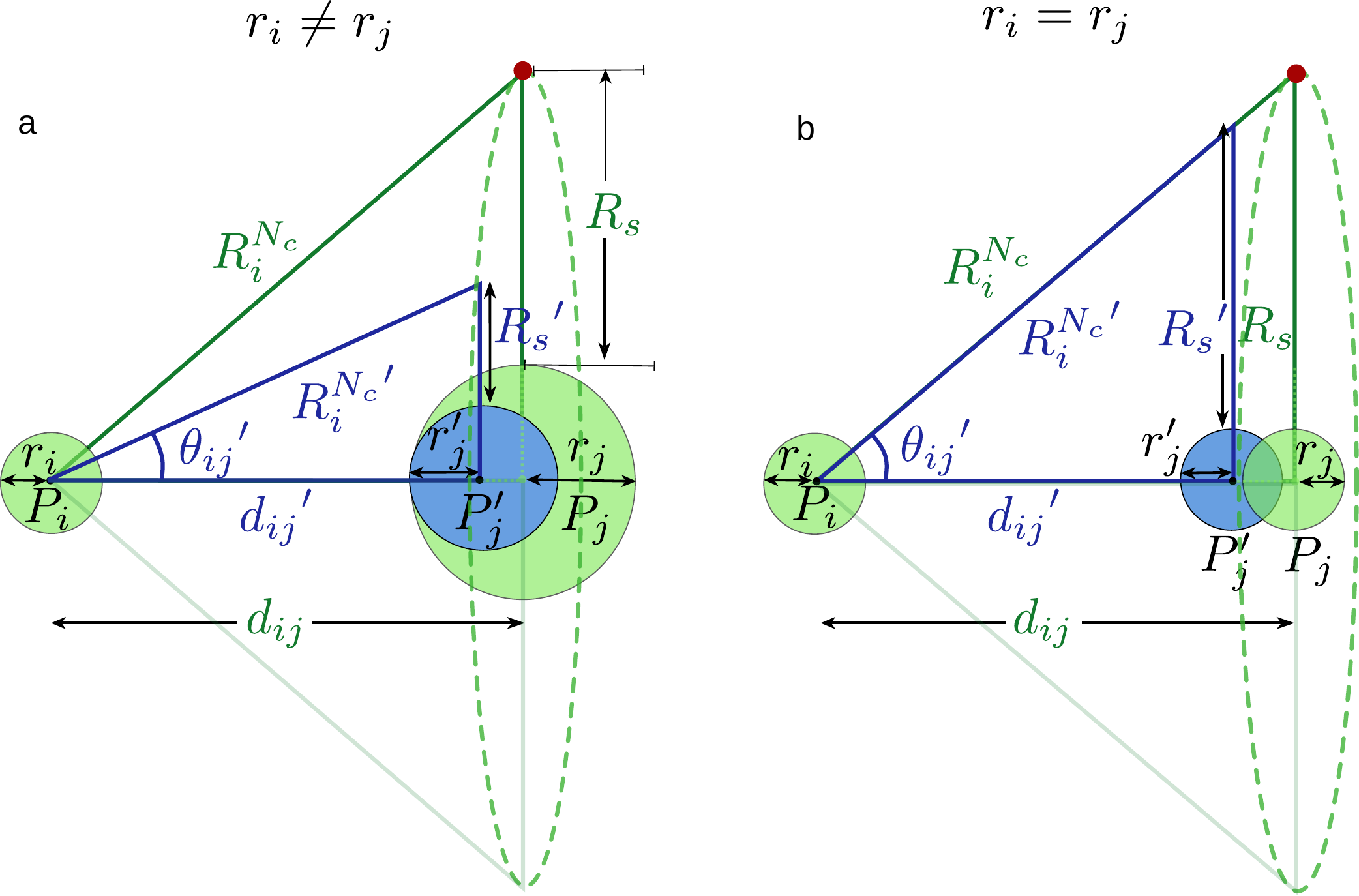}
\caption{
Effect of perturbations on solid-angle blocking in SANN and SANNR. \textbf{a},~Radius rescaling: increasing $r_j$ enlarges the blocked angle (blue to green). \textbf{b},~Displacement: moving $P_j$ closer increases the solid angle. In both cases, SANN overestimates the neighbor count, while SANNR compensates via radius-aware regularization.
}
\label{fig.SANNs}
\end{figure}

This formulation ensures that the blocked solid angle accounts for both position and size. Notably, in the monodisperse limit, the regularized criterion reduces exactly to the original SANN condition, preserving consistency. The effect of radius scaling and particle displacement is illustrated in \cref{fig.SANNs}, which shows how SANNR avoids the overcounting of neighbors that occurs in SANN. The complete computational procedure is summarized in \cref{alg:SANNR}, which defines how the effective blocking radius and coordination number are iteratively determined from geometric and size information.

\begin{figure}
\begin{algorithm}[H]
\caption{SANNR for size-disperse spheres.}
\label{alg:SANNR}
\begin{algorithmic}[1]
\Require Distances $d_{i,j}$ and radii $r_j$ for neighbors of particle $P_i$, sorted by $d_{i,j}$
\Ensure Coordination number $N_c$, coordination radius $R_i^{N_c}$
\For {each particle $P_i$}
	\State Initialize $N_c = 3$
	\State Compute $d_i = d_{i,1} + d_{i,2} + d_{i,3}$
	\While{$N_c < \text{max}$}
		\State $j = N_c + 1$
		\State $d_i \leftarrow d_i + d_{i,j} - (r_i - r_j)$
		\State $R_{i,j}^{\text{tot}} = \sqrt{d_i^2 - (d_{i,j} - (r_i - r_j))^2}$
		\State $R_s = R_{i,j}^{\text{tot}} - r_j$
		\State $d_{\text{tmp}} = \sqrt{1 + \frac{(R_s + r_j)^2}{(d_{i,j} - (r_i - r_j))^2}} \times |d_{i,j} - (r_i - r_j)|$
		\If{$\frac{d_{\text{tmp}}}{N_c - 2} \geq d_{i,j} - (r_i - r_j)$}
			\State $N_c \leftarrow N_c + 1$
		\Else
			\State \textbf{break}
		\EndIf
	\EndWhile
	\State Compute $R_i^{N_c}$ using Eq.~\ref{eq:SANNmainEquation1}
\EndFor
\end{algorithmic}
\end{algorithm}
\end{figure}

\section{Comparison of Neighbor Distributions} \label{results}

We assess the performance of SANNR in both binary and size-disperse mixtures of spheres, using a constant packing fraction of 0.5. Each system contains 5000 spheres, equilibrated using event-driven molecular dynamics (EDMD), following established protocols.\cite{Bommineni2019,Bommineni2019binary} Under these conditions, the system remains in a dense fluid state.

In binary mixtures, particles are divided equally between two species, with a size ratio defined as the radius of the small sphere divided by that of the large one. In size-disperse mixtures, particle radii are sampled from a Gaussian distribution; the standard deviation of this distribution controls the degree of dispersity.

We construct neighbor lists using four different methods: standard Voronoi tessellation (Vr), Laguerre tessellation (Lg), SANN, and SANNR. Voronoi and Laguerre tessellations are computed using the Voro++ software package,\cite{Rycroft2009} and comparisons are made using the Wasserstein distance, a metric that quantifies differences between distributions. All reported values are averaged over ten independent trajectories, with error bars indicating standard errors. Wasserstein distances are computed using the \texttt{SciPy} library.\cite{SciPy2020}

\subsection{Quantifying Neighbor Distribution Similarity} \label{sec.measuredQuantities}

To evaluate how closely the neighbor distributions from SANN and SANNR match those from established tessellation methods, we use the first Wasserstein distance. For two one-dimensional probability distributions $\mu$ and $\nu$, the Wasserstein distance is defined as
\begin{equation}\label{eq:wasserstein_eq}
    W(\mu,\nu) = \inf_{\gamma \in \Gamma(\mu,\nu)} \int_{\mathbb{R} \times \mathbb{R}} |x - y| \, \mathrm{d} \gamma(x, y),
\end{equation}
where $\Gamma(\mu, \nu)$ denotes the set of all joint probability measures with marginals $\mu$ and $\nu$.\cite{Biau2011} In the context of optimal transport, $W(\mu, \nu)$ represents the minimum cost to morph one distribution into the other. It accounts for both shape and location differences and remains stable even when distributions do not overlap.\cite{Xu2018,Goodman1997}

Each neighbor-detection method yields a discrete distribution over the number of nearest neighbors $k$. Let $N_k$ denote the number of particles with exactly $k$ neighbors, and let $N = \sum_k N_k$ be the total particle count. Then, the normalized neighbor distribution is $\mu(k) = N_k / N$. The distribution mean $\bar{N} = \text{E}[\mu]$ represents the average coordination number. For example, in dense Lennard-Jones fluids, Voronoi tessellation typically gives $\bar{N} \approx 14$, whereas SANN gives $\bar{N} \approx 12$ due to the inclusion or exclusion of second-shell neighbors.\cite{vanMeel2012} Similar trends persist in binary mixtures.\cite{Park2012}

\subsection{Binary Mixtures}\label{secsec.binary}

We begin by analyzing binary mixtures with varying size ratios. When the system is monodisperse (size ratio = 1), both Laguerre and SANNR reduce to their radius-insensitive counterparts, Voronoi and SANN, respectively, as expected from their definitions. As the size ratio decreases and the asymmetry between particle sizes increases, differences in neighbor detection become more pronounced.

\begin{figure}
	\includegraphics[width=\columnwidth]{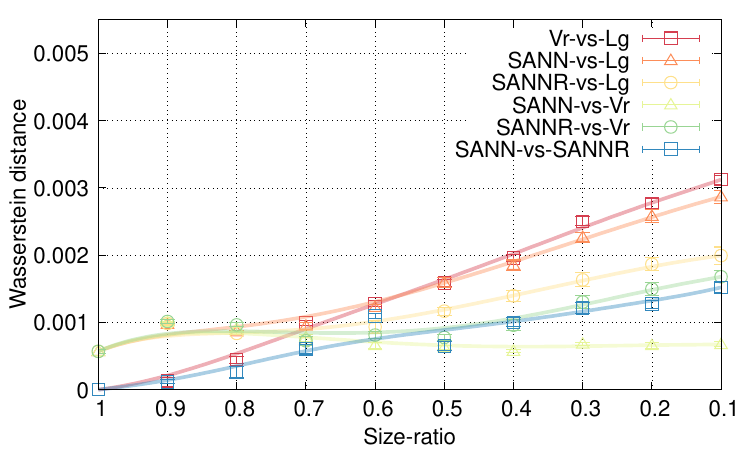}
	\caption{
        Pairwise Wasserstein distances between neighbor distributions from Voronoi (Vr), Laguerre (Lg), SANN, and SANNR in binary mixtures of spheres, shown as a function of size ratio. Lower values indicate greater similarity between methods.
	}
	\label{fig.W43}
\end{figure}

Figure~\ref{fig.W43} shows the pairwise Wasserstein distances between neighbor distributions across the four methods. Radius-independent methods (SANN and Voronoi) maintain a nearly constant distance from each other, indicating insensitivity to particle size. In contrast, radius-sensitive methods (SANNR and Laguerre) diverge more significantly from the radius-insensitive ones as size ratio decreases.

Importantly, SANNR remains closer to Laguerre than to SANN, confirming that it better incorporates particle size. However, the increasing distance between SANNR and Laguerre at smaller size ratios shows that SANNR is not a direct substitute for Laguerre. Rather, it introduces a gradual and regularized incorporation of size effects, providing a distinct and smoother neighbor metric. Meanwhile, the slow divergence between SANN and SANNR shows that SANNR retains compatibility with geometric methods while extending them to systems with moderate size asymmetry.

\subsection{Size-Disperse Mixtures}\label{secsec.polydisperse}

In mixtures with continuous size dispersity, the effects of particle radius become more pronounced. Radius-insensitive methods such as SANN and Voronoi may misclassify neighbors based solely on spatial proximity, which can distort local structure characterization.

\begin{figure}
	\includegraphics[width=\columnwidth]{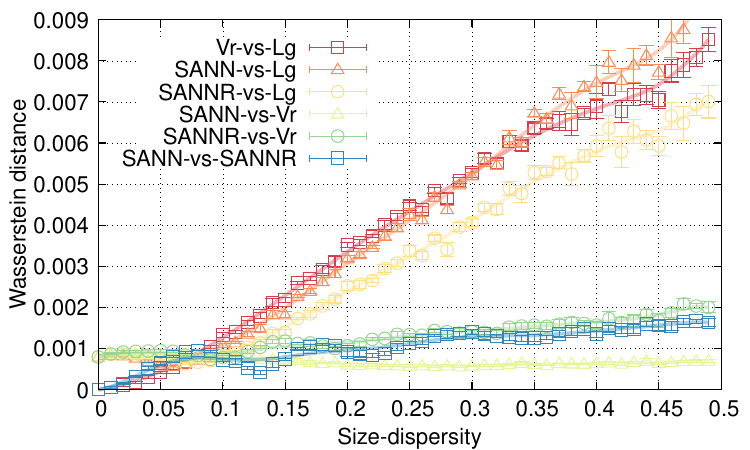}
	\caption{
        Pairwise Wasserstein distances between neighbor distributions from Voronoi (Vr), Laguerre (Lg), SANN, and SANNR in size-disperse mixtures of spheres, plotted as a function of increasing size dispersity. 
	}
	\label{fig.Wasserstein47}
\end{figure}

As shown in \cref{fig.Wasserstein47}, similar trends to the binary case are observed. Across increasing levels of dispersity, the Wasserstein distance between SANNR and Laguerre remains smaller than that between Laguerre and either SANN or Voronoi, again reflecting SANNR’s improved size sensitivity. However, as dispersity grows, SANNR and Laguerre diverge, indicating that they capture size effects differently. This difference underscores that SANNR is not a tessellation-based proxy, but an independent and smoother geometric approach.

Distances between SANN and Voronoi, as well as between SANNR and Voronoi, remain relatively flat, indicating that geometric structure is preserved. Meanwhile, the gradual growth in the distance between SANN and SANNR illustrates how SANNR builds on the geometry of SANN while adapting to the underlying size distribution. These results collectively demonstrate that SANNR offers a reliable extension of SANN for both binary and polydisperse systems, preserving structural intuition while enhancing radius sensitivity.

\section{Application to Order Parameters} \label{secsec.orderParameters}

Order parameters are widely used to characterize structural organization in particle systems and to identify phase transitions. Among the most versatile descriptors are the bond orientational order parameters introduced by Steinhardt and co-workers,\cite{Steinhardt1983} which capture angular correlations in the local environment. These parameters and their many variants\cite{Honeycutt1987, Faken1994, Ackland2006, Stukowski2012, Malins2013, Martelli2018, Bozic2021, tenWolde1995, Auer2004, Lechner2008, Mickel2013, Eslami2018} have proven useful in distinguishing crystalline structures such as face-centered cubic (FCC), hexagonal close-packed (HCP), and body-centered cubic (BCC), as well as for detecting amorphous or liquid-like arrangements.

\subsection{Bond Orientational Order and Local Structure Correlation}

To compute these parameters, one must first define a nearest-neighbor list for each particle, which can be constructed using a fixed cutoff, Voronoi tessellation, or the SANN or SANNR methods introduced earlier. For a particle $P_i$ with $N_i$ neighbors, the local bond orientational order parameters are defined as
\begin{equation}
    q_{lm}^i = \frac{1}{N_i} \sum_{j=1}^{N_i} Y_{l,m}(\hat{\vec{r}}_{ij}),
\end{equation}
where $Y_{l,m}$ are spherical harmonics and $\hat{\vec{r}}_{ij}$ is the unit vector from $P_i$ to $P_j$. These descriptors can be averaged across the system to define
\begin{equation}
    \bar{q}_{lm} = \frac{1}{N} \sum_{i=1}^{N} q_{lm}^i = \frac{1}{N} \sum_{i=1}^{N} \left( \frac{1}{N_i} \sum_{j=1}^{N_i} Y_{l,m}(\hat{\vec{r}}_{ij}) \right),
\end{equation}
from which the global Steinhardt order parameter is obtained as
\begin{equation} \label{global_Steinhardt_OP}
    Q_l = \sqrt{ \frac{4\pi}{2l + 1} \sum_{m=-l}^{l} \left| \bar{q}_{lm} \right|^2 }.
\end{equation}

To quantify similarity between local environments, we compute the normalized correlation coefficient
\begin{equation}
    C_l(q_{lm}^i,q_{lm}^j) = \frac{\sum_{m=-l}^{l} q_{lm}^i q_{lm}^j}{\sqrt{\sum_{m=-l}^{l} |q_{lm}^i|^2} \sqrt{\sum_{m=-l}^{l} |q_{lm}^j|^2}}.
\end{equation}
If rotational invariance is desired, one can maximize this quantity over all 3D rotations,
\begin{equation}
    C_l(i,j) = \max_{R \in \text{SO}(3)} C_l(q_{lm}^i, Rq_{lm}^j),
\end{equation}
where $Rq_{lm}^j$ denotes the bond order vector of particle $P_j$ after applying rotation $R$. This rotationally aligned measure, often referred to as the local–local correlation order parameter, enables unsupervised classification of structurally similar environments.

To assess the impact of neighbor-detection strategy on local order metrics, we examine a simulation snapshot of the AB$_{13}$ crystal, a binary structure containing 104 small and 8 large spheres per unit cell, which belongs to the Frank–Kasper family of phases.\cite{Frank_Kasper_1958,Frank_Kasper_1959} This structure has been observed in molecular dynamics simulations\cite{Bommineni2019binary} and in binary nanocrystal superlattices.\cite{Abbas2008,Ye2011,Ye2015,Coropceanu2019} Its complex local geometry makes it challenging to resolve using conventional order parameter analysis.\cite{Coli2021}

\begin{figure}
	\includegraphics[width=\columnwidth]{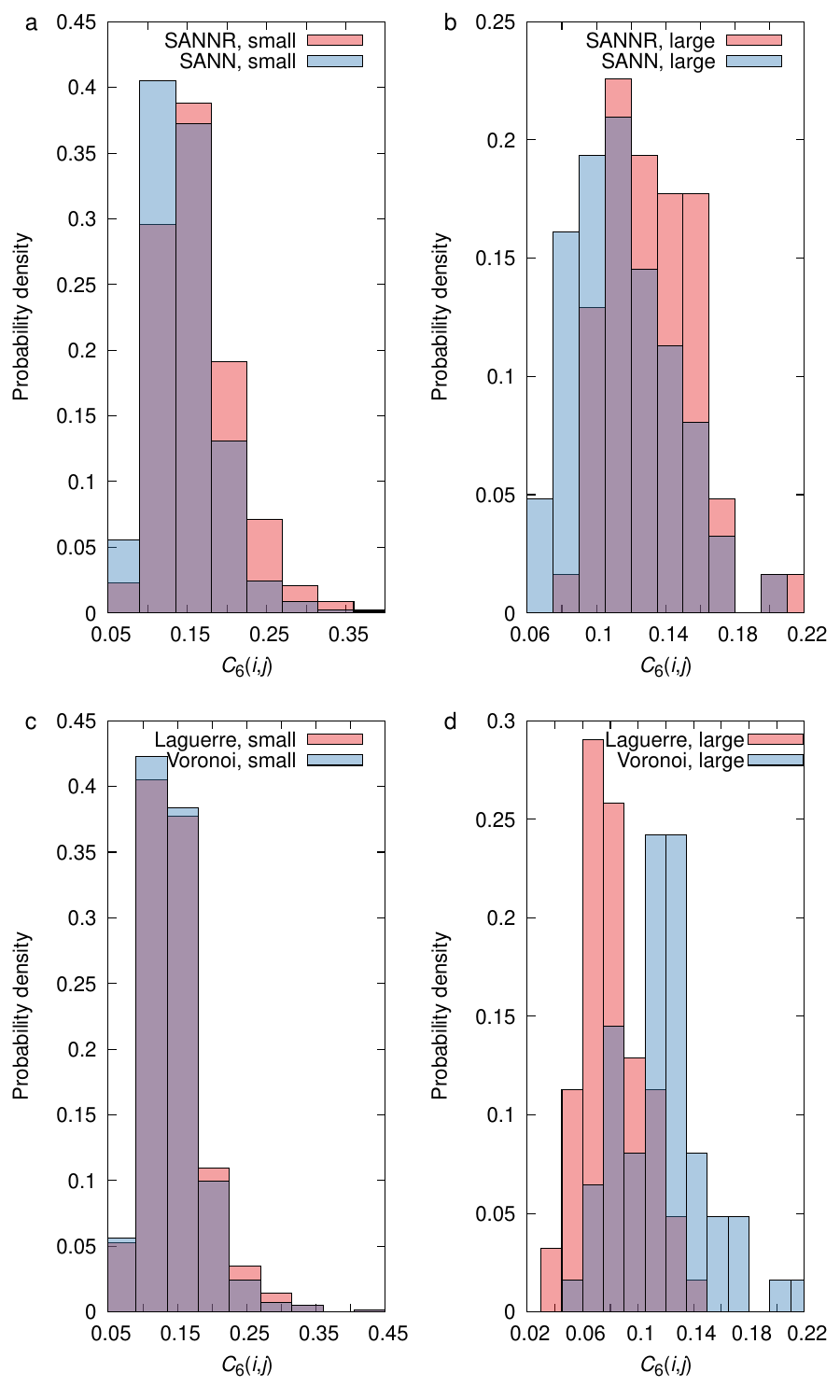}
	\caption{Histograms of $C_6(i,j)$ values for small and large particles in the AB$_{13}$ structure, comparing different nearest-neighbor methods. In \textbf{a}, we compare SANN (blue) and SANNR (red) for small particles and observe that SANNR yields higher $C_6(i,j)$ values. A similar trend appears in \textbf{c}, where Laguerre and Voronoi methods are compared for small particles, again showing that the size-sensitive approach results in higher correlations. All histograms use a bin width of 0.05.}
	\label{fig.applicationssannrclassification}
\end{figure}

We compute $C_6(i,j)$ using neighbor lists generated by SANN, SANNR, Voronoi, and Laguerre tessellations. The rotational alignment is optimized by stochastic gradient descent, and particles are clustered by $C_6$ similarity using the DBSCAN algorithm,\cite{Ester_1996} as implemented in \texttt{scikit-learn}.\cite{scikit-learn} \cref{fig.applicationssannrclassification} shows histograms of $C_6(i,j)$ values for small and large particles. For both species, size-sensitive methods (SANNR and Laguerre) yield higher correlations, indicating that incorporating radius information improves the identification of coherent local environments in the AB$_{13}$ phase.

\subsection{Global Order During Crystallization}

To test the sensitivity of global ordering metrics across time, we analyze a full crystallization trajectory forming AB$_{13}$, as reported in Ref.~\citenum{Bommineni2019binary}. For each frame, we compute the global $Q_6$ order parameter using neighbor lists determined by SANN, SANNR, Voronoi, and Laguerre.

\begin{figure}
	\includegraphics[width=1\columnwidth]{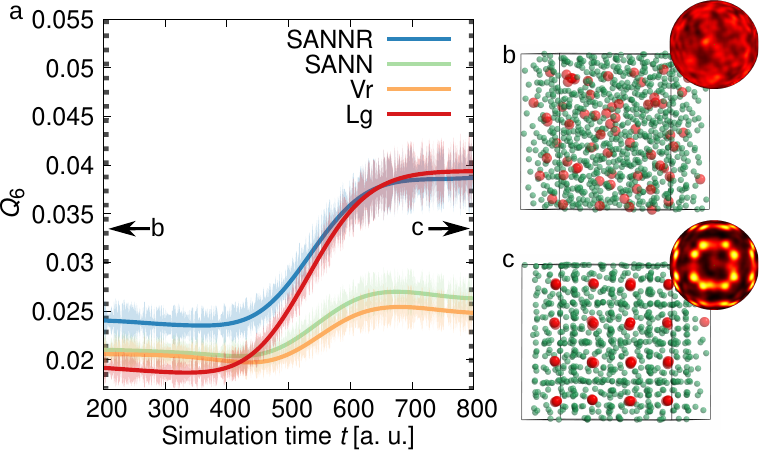}
	\caption{Analysis of a simulation trajectory forming an AB$_{13}$ crystal from the liquid. \textbf{a}, Time evolution of the $Q_6$ order parameter with nearest neighbors detected using SANNR (blue), SANN (green), Voronoi (orange), and Laguerre (red). \textbf{b, c}, Simulation snapshots at the beginning and end of the trajectory. Red and green spheres represent large and small particles, respectively, scaled for visual clarity. Insets show bond-orientational order diagrams: (b) lacks sharp peaks, indicating disorder, while (c) displays peaks consistent with cubic symmetry, confirming crystallinity.}
	\label{fig.AB13_global_Q6_comparison}
\end{figure}

\cref{fig.AB13_global_Q6_comparison}a shows that $Q_6$ increases significantly near simulation time 500, signaling the onset of nucleation. The magnitude of this increase is greater when using SANNR and Laguerre, suggesting that radius-aware methods are more sensitive to the development of orientational order. This is particularly valuable for detecting early-stage structural transitions in polydisperse or complex fluids. Simulation snapshots at the beginning and end of the trajectory (\cref{fig.AB13_global_Q6_comparison}b,c) illustrate the corresponding structural change. The bond-orientational diagram in the disordered liquid shows no discernible features, while the crystalline state exhibits clear peaks consistent with cubic symmetry.

These results confirm that SANNR effectively captures both local and global structural changes, bridging the simplicity of geometric heuristics with the accuracy of tessellation-based methods.

\section{Conclusion}

Nearest-neighbor detection underpins structural classification and phase analysis in particle-based systems. While numerous strategies exist, methods that are both parameter-free and sensitive to particle size remain rare. SANNR contributes to this space by combining the conceptual simplicity of geometric criteria with the ability to account for size disparity, offering a balanced alternative to tessellation-based approaches.

Beyond its demonstrated advantages in model systems, SANNR highlights several broader considerations for structural analysis in soft and condensed matter. Its smooth, local cutoff construction avoids the geometric instabilities that limit some tessellation methods, especially near defects or in heterogeneous environments. At the same time, its reliance on solid-angle constraints lends itself to a physically intuitive interpretation tied to excluded volume, making it readily adaptable to new contexts.

One area of potential development lies in extending SANNR to particles with anisotropic shapes or directional interactions, where local neighborhood definitions may involve orientation or angular constraints. Another promising direction is the integration of SANNR with dynamic analysis, where temporal coherence of neighborhoods can inform studies of nucleation pathways, aging, or active matter dynamics.

Ultimately, neighbor detection is not a solved problem but a context-dependent one. The contribution of SANNR is to expand the toolkit of available methods with a generalizable, smooth, and size-aware approach. Its extensibility and compatibility with a wide range of structural metrics suggest broad utility for future investigations in simulation and experiment alike.

\section*{DATA AVAILABILITY}
Data supporting the findings of this study are available from the corresponding author on reasonable request.

\begin{acknowledgments}
This work was supported by Deutsche Forschungsgemeinschaft (DFG) under grant EN 905/4-1. Scientific support and HPC resources provided by the Erlangen National High Performance Computing Center (NHR@FAU) under the NHR project b168dc are gratefully acknowledged.
\end{acknowledgments}

\bibliography{SANNreg} 

\end{document}